# Transparency-controlled multiple charge transfer in superconducting junctions with local shot-noise scanning tunneling spectroscopy

Yudai Sato[1,2,3,4], Maialen Ortego Larrazabal[1,5], Jian-Feng Ge[1,6], Ingmar Swart[5], Doohee Cho[7], Wolfgang Belzig[8], Juan Carlos Cuevas[9,10], Milan P. Allan[1,2,3,4,*] and Jiasen Niu[1,2,3,4,†]

[1] Leiden Institute of Physics, Leiden University, Niels Bohrweg 2, 2333 CA Leiden, The Netherlands.
[2] Fakultät für Physik, Ludwig-Maximilians-Universität, Schellingstrasse 4, München 80799, Germany.
[3] Munich Center for Quantum Science and Technology (MCQST), München, Germany.
[4] Center for Nano Science (CeNS), Ludwig-Maximilians-University Munich, Munich 80799, Germany.
[5] Debye Institute of Nanomaterials Science, Utrecht University, PO Box 80000, 3508 TA Utrecht, The Netherlands.
[6] Max Planck Institute for Chemical Physics of Solids, 01187 Dresden, Germany.
[7] Department of Physics, Yonsei University, Seoul 03722, Republic of Korea.
[8] Fachbereich Physik, Universität Konstanz, 78457 Konstanz, Germany.
[9] Departamento de Física Teórica de la Materia Condensada, Universidad Autónoma de Madrid, E-28049 Madrid, Spain.
[10] Condensed Matter Physics Center (IFIMAC), Universidad Autónoma de Madrid, E-28049 Madrid, Spain.
*Contact author: milan.allan@lmu.de and †Contact author: niu@physics.leidenuniv.nl

**ABSTRACT**. Charge transport in superconducting junctions at finite voltages is governed by Andreev reflections, including multiple Andreev reflections, which are processes that enable multiple charge transfer, a hallmark that shot noise can directly quantify. Since the effective charge extracted from shot noise measurements varies with the transparency of the junction, systematic control of transparency is essential but experimentally challenging. Here, we present shot noise scanning tunneling microscopy measurements enabled by a newly developed amplifier, allowing access to different transparency regimes. We perform shot noise measurements on Pb(111) with tunable transparency at 2.2 K and observe that the shot noise evolves from a single electron tunneling regime to multiple charge transfer regime as transparency increases. Our results are quantitatively consistent with theoretical simulations of Andreev reflections and multiple Andreev reflections for a single-channel system. These results establish junction transparency as the key parameter governing the evolution of charge transport and demonstrate that noise-STM is a powerful platform for investigating microscopic charge transport mechanisms with controlled junction transparency at the atomic scale.

## I. INTRODUCTION

In superconducting junctions, charge can be transferred in units larger than a single electron via multiple-charge transport processes. For example, in normal metal (N) - insulator (I) - superconductor (S) junctions Andreev reflection (AR) occurs when a bias voltage $V$ lower than the superconducting gap $\Delta$ is applied across the junction. In this process, an electron in the normal electrode cannot enter the superconducting electrode as a single quasiparticle; instead, it is converted into a Cooper pair inside the superconducting region while a hole is reflected, resulting in an effective transfer of charge $2e$ [Fig. 1(a)]. In superconductor–insulator–superconductor (SIS) junctions, the tunneling current can be mediated by multiple Andreev reflections (MAR), a process effectively transferring multiple charges across the junction. A MAR process of order $n$ corresponds to an effective charge transfer of $ne$ [1] and occurs at low tunneling junction transparency ($\tau$) in the voltage range $2\Delta/n < e|V| < 2\Delta/(n-1)$ as shown in Fig. 1(b).

The effective charge transferred across a junction can be directly probed by shot-noise measurements [2]. Shot noise, which arises from the discrete nature of charge carriers, manifests as fluctuations in the electrical current. In the case of single electron tunneling, the noise is Poissonian, with a zero-temperature power spectral density $S_I = 2qI$ in the low transparency limit, where $q = e$ is the carriers' effective charge and $I$ is the average current. In systems with electron correlations or coherent transport [3-10], however, the effective charge of carriers $q$ can deviate from $1e$. Hence, shot noise provides direct information about the effective charge of the tunneling current. It has long been employed in mesoscopic devices, where they enabled the detection of fractional charges in fractional quantum Hall systems [11, 12] and the observation of Andreev reflections (AR) in NIS junctions [7, 13], as well as multiple Andreev reflections (MAR) in SIS mesoscopic junctions [3-6]. More recently, noise spectroscopy combined with scanning tunneling microscopy (noise-STM) has extended these measurements to the atomic-resolution, local tunneling junctions, allowing the detection of AR with spatial resolution [8-10].

The transparency of the tunneling junction plays a crucial role in determining the contribution of the different charge-transfer processes to the transport. AR and MAR processes rely on repeated transmission across the junction and are therefore highly sensitive to the junction transparency [14]. In the low-transparency regime ($\tau \lesssim 0.01$), the probability of an $n$-th order Andreev process scale as $\tau^n$. Under these conditions, single-electron tunneling can become the dominant process, especially in the presence of thermal or quasiparticle lifetime broadening [15, 16]. Such broadening can introduce quasiparticles within the superconducting gap (quasiparticle poisoning), making the $1e$ process more favorable and reducing the measured effective charge to $q \approx e$ even within the superconducting gap [16, 17], where an ideal (M)AR dominated transport would yield $q > e$. Moreover, in the case of typical planar junctions, the large number $N_{ch}$ of low-transparency tunneling channels averages out higher-order processes, effectively washing out any effect beyond single-electron tunneling [18]. On the other hand, in the higher transparency regime ($0.01 \lesssim \tau \ll 1$) MAR process deviates from $q = ne$ expected in lower $\tau$ regime; $q$ becomes larger than $ne$ in the energy range of $2\Delta/n < e|V| < 2\Delta/(n-1)$ as $\tau$ increases because the contributions of higher-order (more than n) MAR processes dominate [19-21].

Despite several experimental studies of MAR-related shot noise [3-6], a broadly tunable and well-controlled single-channel platform has been lacking. While quantitative agreement has been demonstrated in well-defined cases, such as aluminum atomic point contacts [4], the lack of a well-controlled system with tunable transparency has limited systematic shot-noise investigations in the single-channel limit. This motivates our experiments.

## II. METHODS

Scanning tunnelling microscopy (STM) is ideal for this purpose because it provides a well-controlled tunneling junction. The junction transparency can be tuned continuously by changing the tip-sample distance. By combining shot-noise spectroscopy, STM enables access to detailed information about charge dynamics in the tunneling current across an atomically defined junction [22-24]. This technique has been successfully employed to detect the presence of electron pairs in both s-wave and d-wave superconductors [8, 9, 17, 25]. When using STM, we can avoid the complexities associated with multiple transport channels that are typically present in planar junctions. In addition, the clean vacuum barrier intrinsic to STM junctions is free from the chemical and structural disorder of solid-state tunneling barriers. We can also precisely control the junction transparency. This transparency ($\tau$) can be adjusted by tuning the tip-sample distance ($d$). $\tau$ can be estimated by $(G_0 R_{J,N})^{-1}$, where $G_0$ is the conductance quantum $2e^2/h$, $h$ is Plank's constant and $R_{J,N}$ is the normal state junction resistance. In STM measurements, the tip-sample distance is stabilized via the setpoint-current $I_{T,S}$ and set-bias voltage $V_s$ that is normally chosen from values outside superconducting gap, which gives us $R_{J,N} = V_s/I_{T,S}$. A smaller $R_{J,N}$ is obtained by reducing $d$, which increases $\tau$.

Here, we apply noise-STM to superconducting Pb(111) at a temperature of 2.2 K and control the transparencies. First, in the SIN configuration, we observe that the effective charge inside the superconducting gap to gradually increase as $R_J$ decreases, corresponding to an increase in transparency $\tau$. This indicates that at low $\tau$, thermally excited quasiparticles contribute significantly to the tunneling process even within the gap. Next, we measure the shot noise of an SIS junction and detect effective charges larger than $2e$. We observe quantization of the effective charge and an $R_J$-dependence consistent with the MAR process.

## III. RESULTS
### A. SIN junction

First, we use a PtIr tip and Pb sample to form a SIN junction. Figure 2(a) shows the shot noise $S_I$ as a function of bias voltage $V_B$ taken with two typical $R_J = V_B/I_T$, where $V_B$ is bias voltage applied to the sample and $I_T$ is tunneling current. It's clear that shot noise shows transparency dependent behavior. For more results with different $R_J$, see Sec. 3 of the

Supplemental Material [26]. Note that each $S_I$ curve is obtained at fixed $R_J$ with the STM feedback loop active during the $V_B$ sweep (constant-$R_J$ mode). This data acquisition mode is different form the usual method for acquiring d$I$/d$V$ curves in STM, where the feedback loop is turned off after stabilizing tip height (constant-height mode) and then sweep $V_B$ to obtain d$I$/d$V$ signals. For details, see Sec. 1 and Fig. S3 in the Supplemental Material [26]. In constant-height mode, the transparency $\tau$ is constant with $V_B$-sweep. In contrast, in constant-$R_J$ mode, the tip height can change with $V_B$, and therefore, $\tau$ is also $V_B$-dependent Advantages of the constant-$R_J$ mode are that it is less sensitive to mechanical noise and allows to detect the small signals inside the superconducting gap.

From the data shown in Fig.2a, we can extract the effective charge $q = S_I/2e$ as shown in Fig. 2(a). As expected, $q = 1e$ outside the superconducting gap, i.e., for $|eV_B| > \Delta$, where $\Delta = 1.28$ meV obtained by fitting the d$I$/d$V$ spectrum with the Dynes formula [27] (See Sec. 1 in the Supplemental Material [26]). The noise signal starts to increase at a bias voltage of $|eV_B| \sim \Delta$ and saturates around $2e$ for voltages deep inside the gap. The transition between $1e$ and $2e$ becomes sharper for lower $R_J$ (larger $\tau$). The experimentally observed $R_J$ dependence can be explained by our simulations using the Blonder-Tinkham-Klapwijk (BTK) framework, a single channel theoretical model, as shown in Fig. 2(a) and (b) [15, 28, 29]. More details on the simulation are given in Sec. 2 of the Supplemental Material for further details [26]. Figure 2(c) shows how the value of the effective charge inside and outside the gap ($eV_B/\Delta = -0.75$ and -1.5, respectively) depends on $R_J$. The BTK simulations accurately reproduce the experimentally observed behavior. The value of the effective charge does not reach $q = 2e$ for $eV_B/\Delta = -0.75$ due to single particle ($1e$) contributions to the tunnel current, enabled by the finite temperature of the experiment, as well as life-time broadening, in combination with the low transparency. These results demonstrate the importance of the tunneling transparency $\tau$.

### B. SIS junction

Next, we measure the noise $S_I$ in a SIS junction and extract the effective charge $q$. Figure 3(a) shows our constant-height d$I$/d$V$ results on Pb(111) surface with a Pb tip. The superconducting tip prepared by indenting the PtIr tip into the clean Pb(111) surface [30-32]. We fit the results and obtained superconducting gap amplitude $\Delta = 1.32$ meV, where we assume the gap of tip and sample are same, and quasiparticle lifetime broadening $\Gamma = 0.04$ meV. Figure 3(b) and (c) show $S_I$ and constant-$R_J$ d$I$/d$V$ measured at $R_J = 500$ k$\Omega$. Note that the constant-$R_J$ d$I$/d$V$ is obtained together with the noise, which is different from constant-height d$I$/d$V$ (Fig. 3(a)) that is often obtained in STM measurements. For details of constant-$R_J$ d$I$/d$V$, see Fig. S4. We observe peaks at $eV_B = 2\Delta/n$ in the constant-$R_J$ d$I$/d$V$ curve and corresponding jumps in the $S_I$ curve The additional peaks in the constant-$R_J$ d$I$/d$V$ spectra are indicative of MAR processes taking place. As shown in Fig. 3(e), the extracted $q(V_B)$ shows a multi-step behavior, and reaches values $q > 3e$. This is consistent with the occurrence of MAR.

The, $q(V_B)$ curve does not show the clean step-like structure presented in Fig. 1b. This deviation is due to MAR processes of different order occurring simultaneously. First, we have $q < ne$ in $\Delta/n < e|V_B| < 2\Delta/n$, which is obvious, especially for n = 4, where $q$ decreases as $V_B$ decreases. The reduction can be due to the contribution of excited quasiparticles, which allows lower-order MAR processes (n<4). Second, we have noise enhancement with decreasing $V_B$, accompanied by an effective charge $q$ exceeding $2e$ in the range $\Delta < e|V_B| < 2\Delta$. The reason for this is that transparency is large enough for higher-order multiple Andreev reflection processes to contribute significantly. To understand this quantitatively, we extract the bias-dependent transparency by fitting the experimental constant-$R_J$ d$I$/d$V$ (the red curve in Fig.3(c)) using the single-channel MAR theory [33], where all orders of MAR processes are naturally included. The extracted transparency is shown in Fig. 3(d) and it was then used to compute the noise with the help of the theory [19]. The result for the bias-dependent charge obtained by using the extracted transparency, $\Delta$ and $\Gamma$ is shown as a red curve in Fig. 3(e), which is in agreement with the effective charge extracted from the measurements. To obtain information about the dominant transport processes, we utilize the concept of full counting statistics [20, 21] and compute the charge-resolved current contributions due to the different tunneling processes for the parameters extracted from our fits. We show the results in Fig. 3(f). Note that, as expected, MARs dominate the transport for subgap voltages, but the single-quasiparticle tunneling ($I_1$) still has a finite contribution even at voltages below $2\Delta$ because of the quasiparticle broadening and the finite temperature.

To further confirm that the observed noise originates from MAR processes, we perform noise measurements with different $R_J$ as shown in Fig. 4(a). In each region of $2\Delta/n < eV < 2\Delta/(n-1)$, $q$ becomes larger with lower $R_J$, which is also reproduced by our simulation (solid lines). Figure 4(b) shows measured $q$ (filled circles) and simulated (open squares) values of q for different voltages as a function of $R_J$. For higher $R_J$, $q$ is smaller than $ne$, whereas for lower $R_J$, $q$ is enhanced and exceeds $ne$. We conclude that lower $R_J$ suppresses contributions from excited quasiparticles that allows lower order MAR processes and enhances contributions from higher order MAR processes.

## IV. CONCLUSION

In summary, we have demonstrated shot-noise measurements of SIN and SIS junctions with noise-STM in a tunnelling regime, enabling a direct comparison with single-channel theoretical models of Andreev processes. We clearly observe $q = 2e$ in SIN and $q = ne$ in SIS junctions from the multiple charge transport. Our data are in quantitative agreement with single-channel simulations and confirm transparency as a key parameter governing the evolution of charge transport: increasing $\tau$ enhances higher order MAR processes, leading to $q$ exceeding $ne$, while decreasing $\tau$ suppresses $q$ toward $1e$. This highlights the importance of transparency for engineering and probing correlated charge transport.

## ACKNOWLEDGMENTS

This work was supported by the European Research Council (ERC StG SpinMelt and ERC CoG PairNoise) and by a collaboration between The Kavli Foundation, Klaus Tschira Stiftung, and Kevin Wells, as part of the SuperC collaboration. I.S. and M.O.L were supported by the European Research Council (Horizon 2020 "FRACTAL", 865570). J.C.C. thanks the Spanish Ministry of Science and Innovation (contract no. PID2024-157536NB-C22) and the "Maria de Maeztu" Programme for Units of Excellence in R&D (CEX2023-001316-M). W.B. was financially supported by Deutsche Forschungsgemeinschaft (DFG; German Research Foundation) via SFB 1432 (Project No. 425217212) and Project No. 465140728.

[1] T. M. Klapwijk, G. E. Blonder, M. Tinkham, Explanation of subharmonic energy gap structure in superconducting contacts, *Physica B+C* **109–110**, 1657–1664 (1982).

[2] Y. M. Blanter, M. Büttiker, Shot noise in mesoscopic conductors, *Phys. Rep.* **336**, 1-166 (2000).

[3] P. Dieleman, H. G. Bukkems, T. M. Klapwijk, M. Schicke, K. H. Gundlach, Observation of Andreev reflection enhanced shot noise, *Phys. Rev. Lett.* **79**, 3486-3489 (1997).

[4] R. Cron, M. F. Goffman, D. Esteve, C. Urbina, Multiple-charge-quanta shot noise in superconducting atomic contacts, *Phys. Rev. Lett.* **86**, 4104-4107 (2001).

[5] Y. Ronen, Y. Cohen, J. H. Kang, A. Haim, M. T. Rieder, M. Heiblum, D. Mahalu, H. Shtrikman, Charge of a quasiparticle in a superconductor, *Proc. Natl. Acad. Sci. U.S.A.* **113**, 1743-1748 (2016).

[6] T. Hoss, C. Strunk, T. Nussbaumer, R. Huber, U. Staufer, C. Schönenberger, Multiple Andreev reflection and giant excess noise in diffusive superconductor–normal-metal–superconductor junctions, *Phys. Rev. B* **62**, 4079-4085 (2000).

[7] F. Lefloch, C. Hoffmann, M. Sanquer, D. Quirion, Doubled full shot noise in quantum coherent superconductor-semiconductor junctions, *Phys. Rev. Lett.* **90**, 067002 (2003).

[8] K. M. Bastiaans, D. Cho, D. Chatzopoulos, M. Leeuwenhoek, C. Koks, M. P. Allan, Imaging doubled shot noise in a Josephson scanning tunneling microscope, *Phys. Rev. B* **100**, 104506 (2019).

[9] K. M. Bastiaans, D. Chatzopoulos, J.-F. Ge, D. Cho, W. O. Tromp, J. M. van Ruitenbeek, M. H. Fischer, P. J. d. Visser, D. J. Thoen, E. F. C. Driessen, T. M. Klapwijk, M. P. Allan, Direct evidence for Cooper pairing without a spectral gap in a disordered superconductor above $T_c$, *Science* **374**, 608-611 (2021).

[10] U. Thupakula, V. Perrin, A. Palacio-Morales, L. Cario, M. Aprili, P. Simon, F. Massee, Coherent and incoherent tunneling into Yu-Shiba-Rusinov states revealed by atomic scale shot-noise spectroscopy, *Phys. Rev. Lett.* **128**, 247001 (2022).

[11] R. de-Picciotto, M. Reznikov, M. Heiblum, V. Umansky, G. Bunin, D. Mahalu, Direct observation of a fractional charge, *Nature* **389**, 162–164 (1997).

[12] L. Saminadayar, D. C. Glattli, Y. Jin, B. Etienne, Observation of the e/3 fractionally charged Laughlin quasiparticle, *Phys. Rev Lett.* **79**, 2526-2529 (1997).

[13] X. Jehl, M. Sanquer, R. Calemczuk, D. Mailly, Detection of doubled shot noise in short normal-metal/ superconductor junctions, *Nature* **405**, 50–53 (2000).

[14] W.-T. Liao, S. K. Dutta, R. E. Butera, C. J. Lobb, F. C. Wellstood, M. Dreyer, Multiple Andreev reflection effects in asymmetric STM Josephson junctions, arXiv:2601.09889.

[15] G. E. Blonder, M. Tinkham, T. M. Klapwijk, Transition from metallic to tunneling regimes in superconducting microconstrictions: Excess current, charge imbalance, and supercurrent conversion, *Phys. Rev. B* **25**, 4515-4532 (1982).

[16] J.-F. Ge, K. M. Bastiaans, J. Niu, T. Benschop, M. Ortego Larrazabal, M. P. Allan, Direct visualization of quasiparticle concentration around superconducting vortices, *Appl. Phys. Lett.* **125**, 252601 (2024).

[17] J. F. Ge, K. M. Bastiaans, D. Chatzopoulos, D. Cho, W. O. Tromp, T. Benschop, J. Niu, G. Gu, M. P. Allan, Single-electron charge transfer into putative Majorana and trivial modes in individual vortices, *Nat .Commun.*

**14**, 3341 (2023).
[18] J. Niu, K. M. Bastiaans, J. F. Ge, R. Tomar, J. Jesudasan, P. Raychaudhuri, M. Karrer, R. Kleiner, D. Koelle, A. Barbier, E. F. C. Driessen, Y. M. Blanter, M. P. Allan, Why shot noise does not generally detect pairing in mesoscopic superconducting tunnel junctions, *Phys. Rev. Lett.* **132**, 076001 (2024).
[19] J. C. Cuevas, A. Martín-Rodero, A. L. Yeyati, Shot noise and coherent multiple charge transfer in superconducting quantum point contacts, *Phys. Rev. Lett.* **82**, 4086-4089 (1999).
[20] J. C. Cuevas, W. Belzig, Full counting statistics of multiple Andreev reflections, *Phys. Rev. Lett.* **91**, 187001 (2003).
[21] J. C. Cuevas, W. Belzig, dc transport in superconducting point contacts: A full-counting-statistics view, *Phys. Rev. B* **70**, 214512 (2004).
[22] K. M. Bastiaans, T. Benschop, D. Chatzopoulos, D. Cho, Q. Dong, Y. Jin, M. P. Allan, Amplifier for scanning tunneling microscopy at MHz frequencies, *Rev. Sci. Instrum.* **89**, 093709 (2018).
[23] M. Ortego Larrazabal, J. Niu, J.-F. Ge, Y. Sato, J. P. Cuperus, T. Benschop, K. M. Bastiaans, A. Mozes, I. Swart, M. P. Allan, Cryogenic amplifier with high sensitivity and stability for noise-STM, *Rev. Sci. Instrum.* **97**, 023701 (2026).
[24] F. Massee, Q. Dong, A. Cavanna, Y. Jin, M. Aprili, Atomic scale shot-noise using cryogenic MHz circuitry, *Rev. Sci. Instrum.* **89**, 093708 (2018).
[25] J. Niu, M. Ortego Larrazabal, T. Gozlinski, Y. Sato, K. M. Bastiaans, T. Benschop, J.-F. Ge, Y. M. Blanter, G. Gu, I. Swart, M. P. Allan, Equivalence of pseudogap and pairing energy in a cuprate high-temperature superconductor, arXiv:2409.15928.
[26] See Supplemental Material at [url] for the measurement methods (Sec. 1), the simulation in SIN configuration (Sec. 2) and the additional figures for results in SIN junction (Sec. 3).
[27] R. C. Dynes, V. Narayanamurti, J. P. Garno, Direct measurement of quasiparticle-lifetime broadening in a strong-coupled superconductor, *Phys. Rev. Lett.* **41**, 1509-1512 (1978).
[28] M. Octavio, M. Tinkham, G. E. Blonder, T. M. Klapwijk, Subharmonic energy-gap structure in superconducting constrictions, *Phys. Rev. B* **27**, 6739-6746 (1983).
[29] M. P. Anantram, S. Datta, Current fluctuations in mesoscopic systems with Andreev scattering, *Phys. Rev. B* **53**, 16390-16402 (1996).
[30] M. Ruby, B. W. Heinrich, J. I. Pascual, K. J. Franke, Experimental demonstration of a two-band superconducting state for lead using scanning tunneling spectroscopy, *Phys. Rev. Lett.* **114**, 157001 (2015).
[31] K. J. Franke, G. Schulze, J. I. Pascual, Competition of Superconducting Phenomena and Kondo Screening at the Nanoscale, *Science* **332**, 940-944 (2011).
[32] M. T. Randeria, B. E. Feldman, I. K. Drozdov, A. Yazdani, Scanning Josephson spectroscopy on the atomic scale, *Phys. Rev. B* **93**, 161115(R) (2016).
[33] J. C. Cuevas, A. Martín-Rodero, A. L. Yeyati, Hamiltonian approach to the transport properties of superconducting quantum point contacts, *Phys. Rev. B* **54**, 7366-7379 (1996).

**Figures**

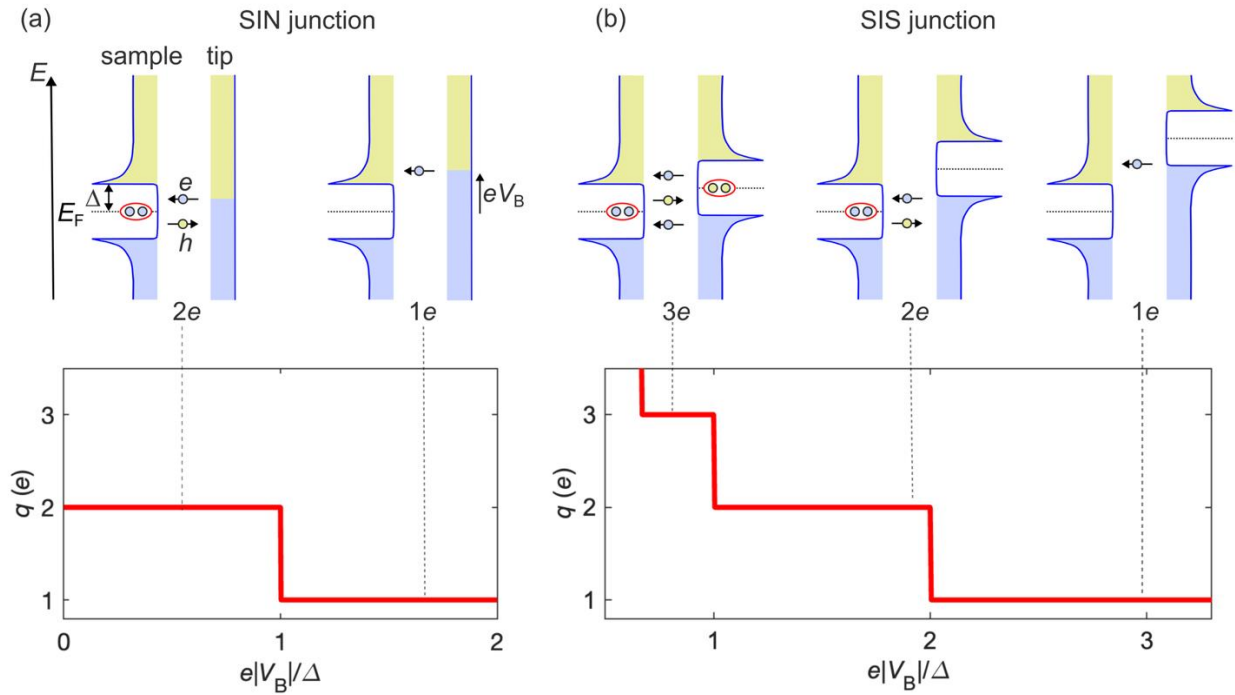

FIG. 1. Schematics of charge transfer in (a) SIN and (b) SIS junctions. The upper panels show sketches of the density of states of two electrodes, while the lower panels show the effective transferred charge q as a function of applied bias voltage $V_B$. (a) In an SIN junction, the normal current at $|eV_B| > \Delta$ is carried by quasiparticles transferring a single-electron charge. For $|eV_B| < \Delta$, Andreev reflection process occurs: An electron transfers a Cooper pair into the superconducting condensate by reflecting a hole in the opposite direction, effectively transferring the 2e charge. (b) In an SIS junction, the tunneling current is dominated by multiple ARs (MAR) of order n, leading to an effective charge transfer of *ne* in the voltage range $2\Delta/n < |eV_B| < 2\Delta/(n-1)$.

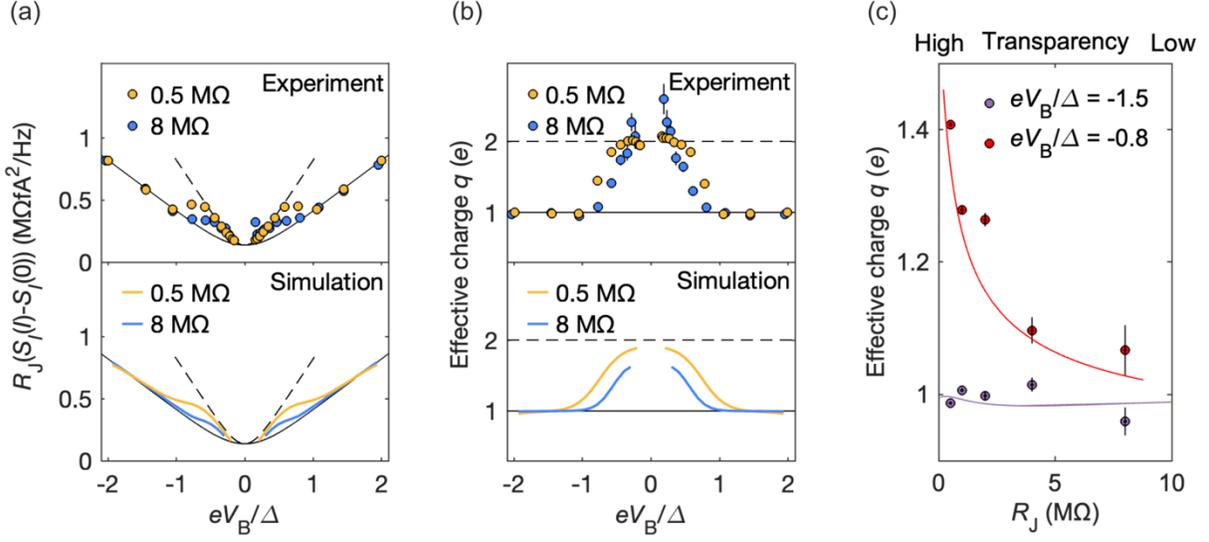

FIG. 2. $R_J$ dependent noise measurements in SIN junction. (a) Measured (top) and simulated (bottom) current noise taken with two typical $R_J$. See Sec. 3 in the Supplemental Material for more results [26]. The solid (dashed) line represents the calculated values for effective charge equals $1e$ ($2e$). (b) Effective transferred charge $q$ extracted from (a). Note that at small bias, the effective charge can be large than $2e$ which could be related to the unstable tip at small bias. (c) Effective transferred charge $q$ at $eV_B/\Delta = -1.5$ and $-0.8$ as a function of $R_J$. The solid lines correspond to simulated values.

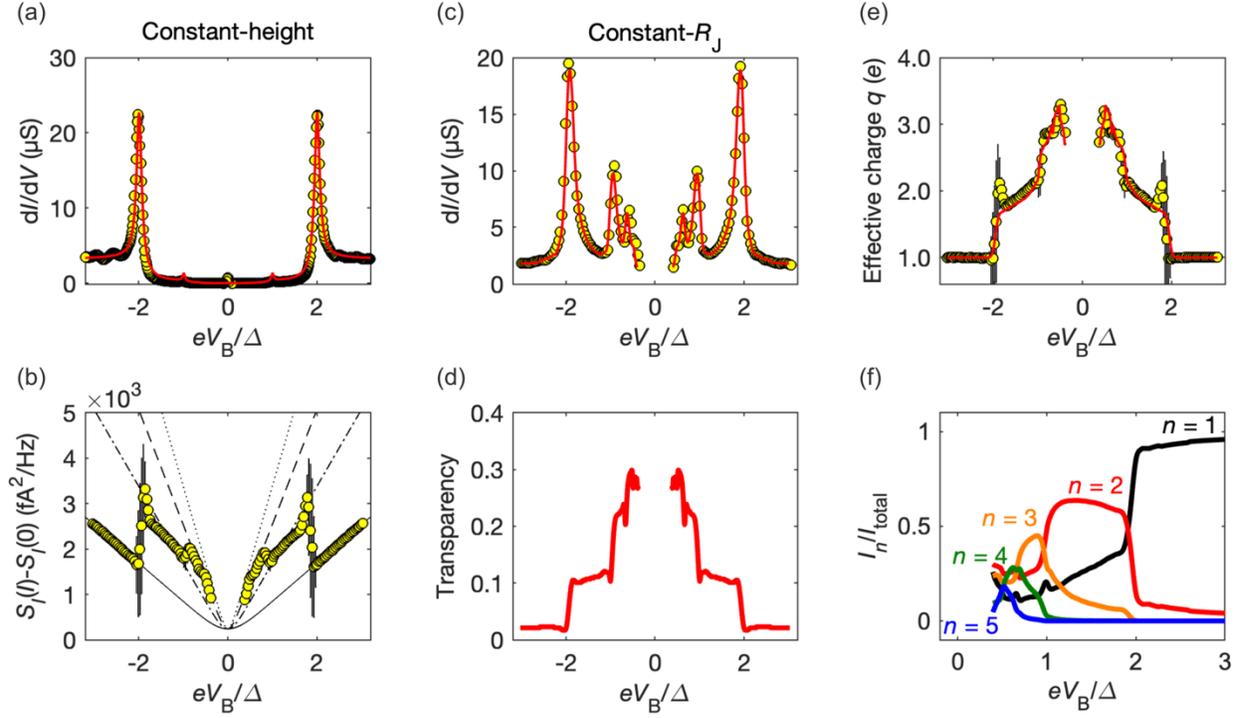

FIG. 3. Noise measurements in SIS junction. (a) Constant-height differential conductance (dI/dV) spectrum. The tip height is stabilized with the tunneling condition of $V_S$ = 5 mV and $I_T$ = 10 nA. The red curve shows a fitted function with $\Delta$ = 1.32 meV and $\Gamma$ = 0.04 meV. (b, c) Measured current noise and d$I$/d$V$ with constant $R_J$ = 500 k$\Omega$. (d) By fitting the constant-$R_J$ d$I$/d$V$ (the red curve in (c)), we can extract transparency as a function of $V_B$. (e) Effective charge $q$ obtain from (b). The red curve is calculated by using $\Gamma$ = 0.04 meV and effective transparency (e). (f) The contribution of $n^{th}$-order MAR with $q=ne$ obtained from the analysis. We use a measurement temperature of $T$ = 2.2 K for all calculations.

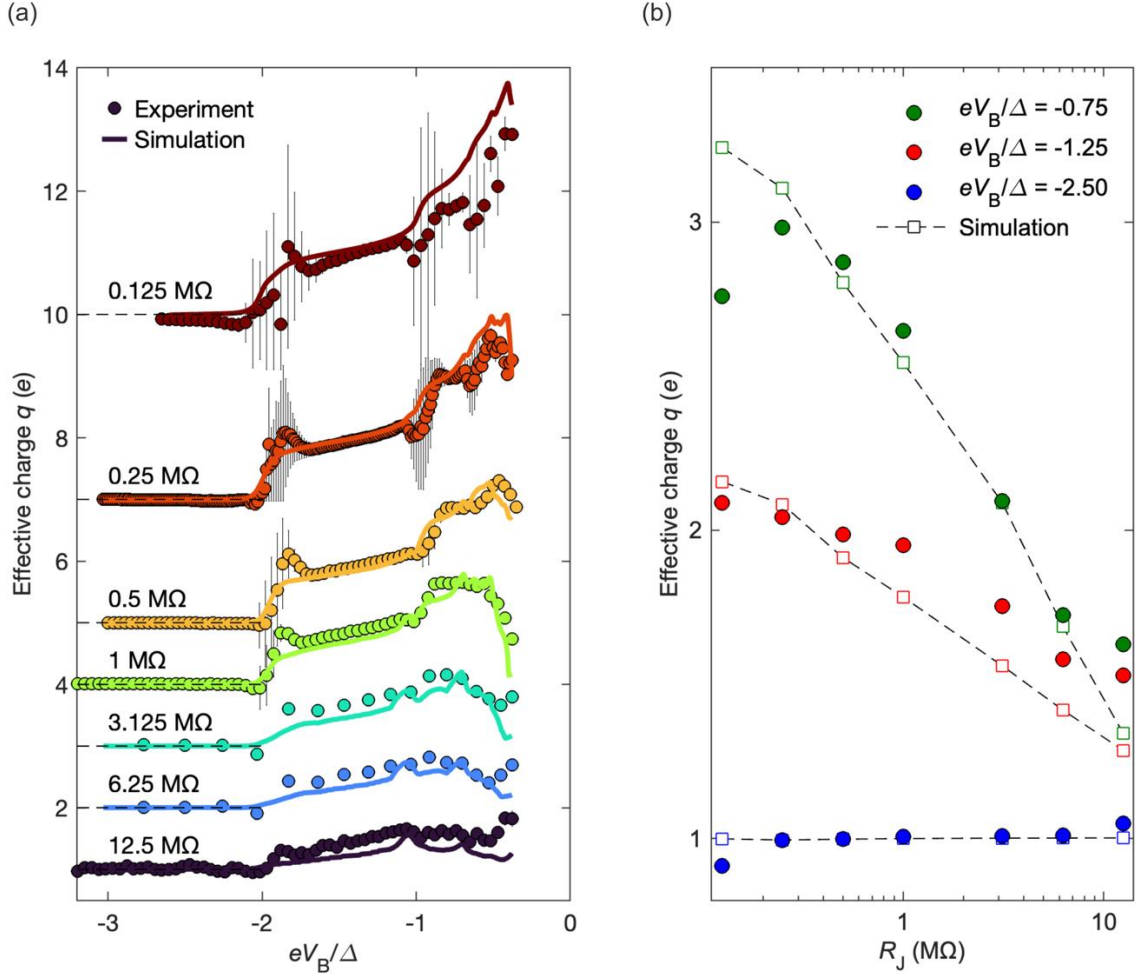

FIG. 4. $R_J$-dependent noise measurements in an SIS junction. (a) Measured (circles) and simulated (solid lines) effective charge $q$ taken with different $R_J$. (b) Dots correspond to the effective charge $q$ at $eV_B = -2.5$ (blue), -1.25 (red) and -0.75 (yellow) as a function of $R_J$. The squares denote simulated values and dashed lines serve as guides to eye.